\documentclass[fleqn,usenatbib]{mnras}

\usepackage{newtxtext,newtxmath}

\usepackage[T1]{fontenc}

\DeclareRobustCommand{\VAN}[3]{#2}
\let\VANthebibliography\thebibliography
\def\thebibliography{\DeclareRobustCommand{\VAN}[3]{##3}\VANthebibliography}

\usepackage{graphicx}	
\usepackage{amsmath}	

\title[GR model of disc galaxies]{Re-weighting dark matter in disc galaxies: a new general relativistic observational test}

\author[D. Astesiano, S.L. Cacciatori, M.Dotti, F.Haardt and F. Re]{
D. Astesiano$^{a,b}$, S.L. Cacciatori $^{a,b}$, M. Dotti$^{c,d}$,  F. Haardt$^{a,d}$ and F. Re$^{a,b,c}$ 
\\
${}^a$ Department of Science and High Technology, Universit\`a dell'Insubria,
Via Valleggio 11, 22100, Como, Italy
\\ 
${}^b$ INFN, sezione di Milano, Via Celoria 16, 20133, Milano, Italy 
\\  ${}^c$ Department of Physics Giuseppe Occhialini, Università di Milano-Bicocca,  Piazza della Scienza 3, 20126, Milano, Italy \\ 
${}^d$ INFN, sezione di Milano-Bicocca,  Piazza della Scienza 3, 20126, Milano, Italy 
}

\date{Accepted XXX. Received YYY; in original form ZZZ}

\pubyear{2022}

\begin{document}
\label{firstpage}
\pagerange{\pageref{firstpage}--\pageref{lastpage}}
\maketitle

\begin{abstract}
 A recent analysis 
 of data from the ESA Gaia mission demonstrated that the kinematics of stars in the Milky Way can be modelled without invoking the presence of dark matter whatsoever. Indeed, the higher-than-Keplerian velocities observed in outer stars can be ascribed to the properties of the general relativistic (GR) metric assumed to describe the Galaxy. Here we generalize the concept, and derive the most general exact GR model for a stationary axi-symmetric  dynamically cold dust structure. 
 We explicitly show how deviations from the commonly adopted Newtonian dynamics are indeed manifest even at low velocities and low densities. We provide for the first time a detailed description of the frequency shift experienced by photons travelling from any emission site within an external disc galaxy to the observer, relating the outcome of the shift measurement to the gravitational properties of the galaxy. Finally, we devise a novel, groundbreaking observational test potentially able to fully characterize the GR metric under the minimal set of assumptions mentioned above. The proposed experiment exploits the effects non-diagonal GR terms have on the frequency shift of photons, ultimately providing a test to evaluate whether dark matter is actually required by disc galaxy kinematics.
\end{abstract}

\begin{keywords}
Galaxy:kinematics and dynamics-- Galaxy:general -- dark matter
\end{keywords}


\section{Introduction and presentation of the results}

The dark matter (DM) hypothesis, i.e., the hypothesis of the existence of a non-baryonic component of mass dominating the matter density budget in the Universe, is one of the foundations of the widely accepted cosmological model \citep[e.g.][]{planck2018b} and, at the same time, one of the greatest unknown in physics. The DM hypothesis has been incredibly successful in interpreting different astrophysical observables, such as the velocity distribution of galaxies in galaxy clusters
\citep[GCs,, since][]{zwicky}, the rotation curves (RCs) of disc galaxies \citep[e.g.][]{rotation_curves}, the thermodynamical properties of X-ray emitting gas in GCs \citep{gas} and the gravitational lensing produced by their mass distributions \citep{lensing}, as well as cosmological ones, e.g., the anisotropies observed in the cosmic microwave background \citep{CMB1, CMB2} and the growth of cosmic structures from such anisotropies \citep{perturb1, perturb2}. However, despite such remarkable successes, results of the many enterprises aimed at the direct detection of DM particles are, to date, inconclusive  \citep[e.g.,][]{reviewdm, directdm1, directdm2}. 

Because of the DM relevance in our understanding of the laws of nature, it is then of paramount importance to fully characterise such elusive component of our Universe, making sure to maximise the outcome of the available indirect astrophysical and cosmological tests. Indeed, the characterization of the nature of DM through the study of the large scale distribution of galaxies is one of the scientific drivers of the medium-class ESA mission Euclid \citep{euclid1, euclid2}.     
With the present study we demonstrate that a fully general relativistic (GR) description of individual disc galaxies can also be used to improve considerably our estimates of their DM content. 

As mentioned above, the current estimates of the DM content in disc galaxies are based on the observed rotation curves, whose flatness at large radii cannot be explained by the distribution of visible (baryonic) matter alone. This assertion, however, relies on a purely Newtonian description of the system. The simplifying assumption of Newtonian gravity is commonly justified by the sub-relativistic velocity of the stars in galaxies, and by the weak gravitational potential involved.

It has been noticed, however, that even in these limits GR corrections may be not-negligible, and, as a consequence, relativistic corrections to Newtonian dynamics \citep{Ruggiero:2021lpf,  Ludwig}, as well as an exact GR model \citep{Balasin:2006cg} have been proposed. All these models are static and axi-symmetric. The exact model by \citet{Balasin:2006cg} has additional simplifying assumptions, considering 
 only dust with zero velocity dispersion, zero mean velocity in the radial and vertical direction and null deformation tensor ($H=-1$ in the following, see section~2 and \citealt{Astesiano:2021ren}). Despite these additional assumptions, the model has been shown to reproduce the kinematic data obtained by Gaia for stars in the Milky Way without invoking any additional component other than visible matter \citep{crosta2020}.

In the present work we continue the generalization of the \citet{Balasin:2006cg} model started in \citet{Astesiano:2021ren}. While preserving the assumption of axi-symmetry, we relax the assumption of null deformation tensor, and derive, for the first time, the frequency shift of a generic spectral line emitted in our GR galaxy model. Making use of such observable, we devise an observational test that could, in principle, fully characterize the metric of any axi-symmetric disc galaxy, therefore providing an accurate estimate of the DM content needed to fit radial velocity profiles.

The outline of the work is as follows: in \S\ref{S2}, we describe the GR model for disc galaxy assuming a stationary and axisymmetric metric filled with dust \footnote{With the term “dust" we refer here to the GR definition, i.e., the energy momentum tensor of the matter is given by (\ref{dust tensor}), with zero pressure and zero stress.}; in \S\ref{S3} we provide general formulae for the description of the motion of the dust; in \S\ref{S4}, we calculate the frequency shift that a photon undergoes from the emission in the galaxy until its observation by an asymptotic inertial observer, using the eikonal approximation; the main observational test is presented in \S\ref{S5}; finally, our conclusions are presented in \S\ref{conclusions}. The reader uninterested in the actual derivation of the model can go directly to \S\ref{S5}, as the mathematical details of the GR solutions in sections \S\ref{S2}, \S\ref{S3} and \S\ref{S4} are not necessary to understand the logic behind the observational test we are proposing.

\section{Preliminary set up \label{S2}}
In the derivation of the metric any internal symmetry-breaking structure, such as spirals or stellar bars, as well as any component with significant velocity dispersion  such as, e.g., stellar bulges, are neglected in favor of the analytic results. Moreover, we consider the velocity dispersion in the disc to be negligible compared to the main global rotational motion. Unlike other general-relativistic galaxy models in the literature \citep[e.g.,][]{Balasin:2006cg, Cooperstock:2005qw} we do not assume a rigidly rotating dust; conversely, we consider the most general differential rotation which is allowed by stationarity and axisymmetry. 

It is well known \citep{HansenWinicour1, Misner:1974qy} that a stationary, axisymmetric metric can be written in the form
\begin{align}
\label{metric}
   ds^2=& -e^{2U(r,z)} \left(dt+A(r,z)d\phi\right)^2 + \\+& e^{-2U(r,z)}  \left[r^2 d\phi^2+ e^{2k(r,z)}\left(dr^2+dz^2\right)  \right]. 
\end{align}
Indeed, stationarity and axial symmetry ensure that it is possible to choose local coordinates such that the coefficients of the metric are independent from the time coordinate $t$ and from the angle $\phi$ of rotation around the symmetry axis. Non staticity implies that in such coordinates the metric is non diagonal, with a $2 g_{t\phi}(r,z)dt d\phi$ term. After a suitable choice of the coordinates the elements of the metric can be then expressed in terms of three arbitrary functions of $r$ and $z$. These are the functions $A(r,z)$ which describes the rotational properties (i.e., non staticity), $U(r,z)$ which essentially measures the time lapses, and $k(r,z)$ which is related to spatial scales.

As in \citet{stephani_kramer_maccallum_hoenselaers_herlt_2003}, we consider when gravity is coupled to dust through the stress-energy tensor 
\begin{align}
\label{dust tensor}
    T^{\mu\nu}= \rho (r,z) u^{\mu} u^{\nu},
\end{align}
where the mass density $\rho$ and four velocity $u$ are 
\begin{align}
    u= \frac{1}{\sqrt{-H}}\left(\partial_t+\Omega(r,z)\partial_{\phi}\right). \label{DV}
\end{align}
Here $\Omega$ represents the angular velocity of the dust with respect to coordinates $t, \phi$. The function $H$ is defined by the condition\footnote{Note that $H$ is defined as negative, consistent with the notation used in  \citet{stephani_kramer_maccallum_hoenselaers_herlt_2003}}
\begin{align}
-1=g_{\mu\nu}u^\mu u^\nu=-\frac 1H (g_{tt} +\Omega^2 g_{\phi\phi} +2g_{t\phi}\Omega). 
\end{align}
Therefore,
\begin{align}
H=-e^{2U(r,z)} \left(1+A(r,z)\Omega(r,z)\right)^2+e^{-2U(r,z)} r^2 \Omega(r,z)^2 \label{Acca}
\end{align}
is a well defined combination of the coefficients of the metric.

The equations of motion are completely integrable and depend on the choice of arbitrary solutions of the equations of motion, as shown below.
Therefore the Einstein equations do not fix $H$ completely, rather they imply that $H$ must be an a priory arbitrary function of another function $\eta(r,z)$.
The importance of passing through $\eta$ is that it has a direct physical interpretation, as we will see in (\ref{velocity}). 

Once the negative function $H(\eta)$ is chosen, we introduce the auxiliary function\footnote{For any function of one variable, like $H(\eta)$, we use a prime sign for the derivative w.r.t. the variable.}
\begin{equation}
\label{def v}
	\mathcal{F}(r,z)=2\eta+r^2\int\frac{H'}{H}\frac{d\eta}{\eta}-\int\frac{H'}{H}\eta d\eta,
\end{equation}
so then $\eta$ is set by the following equation:
\begin{equation}
\label{harm}
	\mathcal{F}_{,rr}-\frac{1}{r}\mathcal{F}_{,r}+\mathcal{F}_{,zz}=0,
\end{equation}
where comma sign indicates partial derivative w.r.t. the specified coordinate. 
In a practical procedure, it is convenient to solve first the latter equation for $\mathcal F$ and then use (\ref{def v}) as a functional equation to determine $\eta$ and then $H$. 

Once $H(\eta)$ and the solution to eq.(\ref{harm}) are known, we are in the position to determine the metric components:
\begin{align}
	g_{tt}&=\frac{(H-\eta\Omega)^2-r^2\Omega^2}{H}, \cr
	g_{t\phi}&=\frac{r^2-\eta^2}{H}\Omega+\eta, \cr
	g_{\phi\phi}&=\frac{r^2-\eta^2}{(-H)} \label{components}, \\
	\Omega&=\frac{1}{2}\int H'\frac{d\eta}{\eta} \label{eta cond}.
\end{align}
We remark that $\Omega$ is determined by an indefinite integral. Therefore $H$ equal to a constant would imply a constant $\Omega$, non necessarily equal to zero, .
Finally, the remaining metric components are
\begin{align}
g_{zz}=g_{rr}=e^{2(k-U)}\equiv e^{\mu}, 
\end{align}
where the function $\mu$ is determined by
\begin{align}
    \mu_{,r} =& \frac{1}{2r} \left[ (g_{tt})_{,r} (g_{\phi\phi})_{,r}-(g_{tt})_{,z} (g_{\phi\phi})_{,z} - ((g_{t\phi})_{,r} )^2+((g_{t\phi})_{,z} )^2 \right]     \label{mur},\\
    \mu_{,z}= & \frac{1}{2r} \left[ (g_{tt})_{,z} (g_{\phi\phi})_{,r} + (g_{tt})_{,r} (g_{\phi\phi})_{,z} -  2 (g_{t\phi})_{,r} (g_{t\phi})_{,z}  \right] \label{muz}.
\end{align}
The role played by the function $\eta$ is clear when we consider that the tri-velocity of the dust $v$ as measured from Zero Angular Momentum Observers (ZAMOs\footnote{ZAMOs, consistently, satisfy the requirement of zero angular momentum, i.e., $g_{\phi\phi}\, \chi+g_{\phi t}=0$.}, 
see section \ref{S3}) results to be 
\begin{align}
\label{velocity}
    v(r,z)=\frac{\eta(r,z)}{r},
\end{align}
in natural units (so that the speed of light is $c=1$). For this reason it is also convenient to rewrite the constraint $(\ref{harm})$ directly as a functional equation for the velocity $v(r,z)$, i.e., 
\begin{align}
\label{eq0}
    0=&l'\left(\frac{1}{v}-v\right)[r^2v_{,z}^2+(rv_{,r}+v)^2]+\cr
    &+l \left[\left(\frac{1}{v}-v\right)(rv_{,zz}+rv_{,rr}+3v_{,r})-\left(\frac{1}{v}+v\right)\frac{r}{v}(v_{,z}^2+v_{,r}^2)+\frac{2}{r}\right]+\cr
    &+2\left(v_{,zz}+v_{,rr}+\frac{v_{,r}}{r}-\frac{v}{r^2}\right),  
\end{align}
where $l= H'/H$ and $H'= dH/d\eta$. 

In general, $H$ can be assumed a priory, then eq.~\ref{eq0} is a differential equation for the field $v(r,z)$. However, in our context it would be relevant to determine $H$ through, e.g., comparison with observational data. In this case, the above eq.~\ref{eq0} becomes a (rather cumbersome) functional equation.

Finally, the matter density is
\begin{equation}
\label{rho}
	8\pi G\rho=\left(\frac{v^2(2-\eta l)^2-r^2l^2}{4g_{rr}}\right)\left(\frac{\eta_{,r}^2+\eta_{,z}^2}{\eta^2}\right),
\end{equation}
and the metric components satisfy the following constraints:
\begin{align}
	-\det(g_{ij})_{i,j=t,\phi}:=g_{t\phi}^2-g_{tt}g_{\phi\phi}&=r^2. \label{det}
\end{align}


\section{Considerations on the assumed metric and on the angular momentum of test particles \label{S3}}
The assumed metric (eq.~\ref{metric}) can be recasted into the following form,
\begin{align}
    ds^2&=\tilde g_{\mu\nu} dx^\mu dx^\nu =- e^{2\nu} dt^2 + e^{2\psi}\left(d\phi-\chi dt \right)^2+ e^{\mu}\left(dr^2+dz^2\right), 
\end{align}
where
\begin{align}
    e^{2\nu}&=\frac{r^2}{g_{\phi\phi}}, \quad e^{2\psi}=g_{\phi\phi},\quad \chi \equiv - \frac{g_{t\phi}}{g_{\phi\phi}}=\Omega+\frac{H \eta}{(r^2-\eta^2)} \label{IR}. 
\end{align}
allowing us to define a tetrad corresponding to ZAMOs, i.e.,
\begin{gather}
    e^0= \frac{r}{\sqrt{g_{\phi\phi}}} dt, \qquad  e^1=\sqrt{g_{\phi\phi}} (d\phi-\chi dt), \\\qquad e^2= e^{\mu/2}\, dr,\qquad e^3= e^{\mu/2}\, dz.
\end{gather}
The congruence of the wordlines associated to ZAMOs carries an angular velocity $\chi=\chi(r,z)$ as measured from a Minkowskian observer at infinity. Equation (\ref{IR}) combined with (\ref{velocity}) can be rewritten in the more meaningful way
\begin{align}
    r (\Omega - \chi)= -H \frac v{1-v^2}.  \label{Chi}
\end{align}
The above expression states that $v$ is a measure of the rotation velocity, $\omega =\Omega-\chi$ represents the relative angular velocity respect to ``non rotating'' ZAMOs, so, for weak field ($H\sim -1$) and non relativistic velocities, it takes the form $r\omega=v$. As an example, in the rigidly rotating model by \citet{Balasin:2006cg}, defined by $H=-1$ and $\Omega=\Omega_0$ as constant, eq.~(\ref{Chi}) implies that the true physical motion (relative to ZAMOs) is given by $v(r,z)$. It follows that the ``gravitomagnetic'' term $\chi$ is
\begin{align}
\chi(r,z)= \Omega_0 - \frac{v(r,z)}{r} \frac 1{1-v^2(r,z)} = \Omega_0 - \frac{v(r,z)}{r} + O(v^2).
\end{align}
Then, although the angular velocity measured from an asymptotic inertial observer $\Omega_0$ is constant, $v(r,z)$ has a non trivial profile and, consistently with GR, the motion of matter is not rigid in a strict sense.

Let us now consider the motion of a test mass having four velocity 
\begin{align}
    \vartheta= \vartheta^t \partial_t + \vartheta^r \partial_r + \vartheta^z \partial_z +\vartheta^{\phi} \partial_{\phi}.
\end{align}
In the given background there are two conserved quantities for such motion, i.e., 
\begin{align}
    E=-g_{0\mu} \vartheta^\mu,\quad M=g_{\phi\mu} \vartheta^\mu,
\end{align}
clearly corresponding to the conservation of energy and of the $z$-component of the angular momentum.
From their expressions, we can then write
\begin{align}
    \vartheta^t=& \frac{E g_{\phi\phi}+Lg_{t\phi}}{g_{t\phi}^2-g_{tt}g_{\phi\phi}}= \frac{g_{\phi\phi}}{r^2} \left(E-\chi M\right), \\
        \vartheta^\phi=& -\frac{E g_{t\phi}+Lg_{tt}}{g_{t\phi}^2-g_{tt}g_{\phi\phi}}=\frac{g_{\phi\phi}}{r^2} \left(E-\chi M\right) \chi + \frac{M}{g_{\phi\phi}},
\end{align}
which, from $\Omega=\vartheta^\phi/\vartheta^t$, allows us to write the physical angular velocity $\omega$ as
\begin{align}
    \omega= \Omega- \chi = \frac{H^2}{r^2 (1-v^2)^2} \frac{M}{(E-\chi M)}. \label{Angularvelocity}
\end{align}
The condition $g(\vartheta,\vartheta)=-1$ implies
\begin{align}
   g_{zz} (\vartheta^z)^2+ g_{rr} (\vartheta^r)^2+ 1+ \frac{M}{g_{\phi\phi}} - \frac{g_{\phi\phi}}{r^2}. (E-M\chi)^2=0,\label{EC}
\end{align}
The above equation would be the same for the motion of a particle in an effective potential $V_{\text{eff}}$ given by \citep[see][]{Herrera-Aguilar:2012jhe}
\begin{align}
    2V_{\text{eff}}=- 1+ \frac{H M^2}{r^2(1-v^2)} - \frac{1-v^2}{H} (E-M\chi)^2.
\end{align}
If we are interested in circular orbits, we have to impose 
\begin{align}
      \partial_r V_{\text{eff}}=0,  \qquad \partial_z V_{\text{eff}}=0, \label{romega-chi}
\end{align}
which, because of (\ref{EC}), gives
\begin{align}
 V_{\text{eff}}=0,
\end{align}
hence relating $E$ and $M$ for the given circular motion. The values of the angular velocity along circular orbits is therefore
\begin{align}
    \omega= \frac{rM}{g_{\phi\phi}}\frac{1}{\sqrt{M^2+g_{\phi \phi}}}.
\end{align}
As expected, in case $M=0$ we have a unique orbit which represent the trajectory of the ZAMO. Replacing the relation between $\omega$ and $v$ we obtain the following expression for $v$,
\begin{align}
    v= \pm \frac{1}{\sqrt{1+ \frac{g_{\phi\phi}}{M^2}}}, 
\end{align}
which relates it to the geometry ($g_{\phi\phi}$) and the angular momentum.

\section{Frequency shift of light from disc galaxies \label{S4}}
We will now determine the frequency shift of a photon emitted by a particle of dust, as detected by asymptotically inertial detectors. The emitters are supposed to move in stable circular geodesic motions.\footnote{This implies that our approximations are valid only in the regime where the velocity dispersion is negligible with respect to the mean motion of the stars. If the dispersion is not too large, then its effect can be simply then taken into account by estimating the kinematic Doppler effect, without the need to account for further gravitational corrections.} We will assume that, after emission, the photons propagate along null geodesics, thus neglecting any possible relativistic aberration. Our results are essentially a generalization of what is presented in \citet{Herrera-Aguilar:2012jhe}. 
In general, the frequency shift is the combination of gravitational and Doppler effects. After the removal of the galaxy peculiar motion and Hubble expansion, the observed shift is usually interpreted in pure kinematic terms. However, we will show that in a genuine general relativistic galaxy models the gravitational frequency shift can be of the same order of the kinematic Doppler effect. 

Let's start by considering a disc galaxy with the galactic plane tilted by an angle $\lambda$ with respect to the line of sight. We assume conventionally $\lambda>0$ when the galaxy appears rotating counterclockwise. We neglect the relativistic aberration, since we are assuming that the gravitational fields are indeed weak. Such an assumption allows us to work in eikonal approximation
\citet{Misner:1974qy}. We also assume that Hubble expansion and any possible galaxy peculiar motion can be removed from the observed shift (see Section \ref{z est}). Moreover, we assume that the detector is at rest in an asymptotically flat region far away from the galaxy, which is assumed to be isolated, so that the four-velocity of the detector is $\partial_t$ in the coordinates of (\ref{metric}). Since, in a real context, observations are made from inside the Milky Way, to consistently compare our model to observations one should also remove the effects associated to the local properties of the Galaxy metric. Because of to the small angular size of the observed galaxy, that ensures that all collected photons travelled through similar paths, such correction can be treated as a ``systematic shift".

Now, assume that the photon is emitted at frequency $\omega_e$ by a dust particle, and observed at  frequency $\omega_d$, with $1+z=\omega_e/\omega_d$ the observed redshift. The proper frequencies are
\begin{equation}
\label{energie}
	\omega_{e,d}:=-U^{\mu}_{e,d}k_{\mu},
\end{equation}%
where $U^{\mu}_d$ is the four-velocity of the detector, $U^{\mu}_e$ is the four-velocity of the source, and $k_{\mu}$ is the photon's four-momentum along its world line.
A photon moving freely on a curved background is described by a gauge field $A^\nu(x^{\mu})$ obeying, in the Landau-Lorentz gauge, the wave equation
\begin{equation}
	\Box A^\mu= R^\mu_\nu A^\mu,
\end{equation}
where $\Box$ is the D'Alembertian acting on vector fields and $R^\mu_\nu$ is the Ricci tensor. As shown in \citet{Misner:1974qy}, for wavelengths much smaller than the scale length of the spatial curvature, these equations can be solved in the eikonal approximation%
\begin{equation*}
	\Psi(x^{\mu})=\Psi_0 e^{i\varphi} \quad s.t. \quad d\varphi=k_{\mu}dx^{\mu},
\end{equation*}%
where we recall that the four-momentum of the photon satisfies $k^{\mu}k_{\mu}=0.$ The time dependence of the phase is only through the term $k_t x^0=-\omega t$, where $\omega:=-k_t$ is the energy of the photon (relative to the coordinate time), which is conserved along the photon's trajectory. For the spatial components of the four-momentum one defines an ``effective refraction index'' $\alpha$ such that
\begin{equation}
	|\vec{k}|=:\alpha\omega,
\end{equation}
where $|\vec k|$ is defined by
\begin{equation*}
	|\vec k|^2=g^{\phi\phi}k_{\phi}^2+g^{rr}k_r^2+g^{zz}k_z^2=-\frac{g_{tt}}{r^2}k_{\phi}^2+e^{-\mu}(k_r^2+k_z^2),
\end{equation*}
It is also convenient to use $\phi$ and $\lambda$ as spherical coordinates with azimuthal angle $\lambda$, so that 
\begin{eqnarray}
    &k_{\phi}=\omega\alpha\frac{r}{\sqrt{-g_{tt}}}\sin\phi\cos\lambda, \cr
    &k_r=\omega\alpha e^{\mu/2}\cos\phi\cos\lambda, \cr
    &k_z=\omega\alpha e^{\mu/2}\sin\lambda.
\end{eqnarray}
Finally, we introduce the angle $\theta$ of projection along the direction $\partial_\phi$, such that 
\begin{align}
\sin\theta=\sin\phi\cos\lambda. 
\end{align}


We can now determine the frequency shift $z$. Since the detector four-velocity is $U^{\mu}_d=\partial_t$ the detected proper frequency is $\omega_d\equiv \omega$.
On the other hand, the four-velocity of the emitter is given by eq,~(\ref{DV}), so we get
\begin{align}
\label{redshift}
	1+z&=\frac{1}{\sqrt{-H}}\left(1-\Omega\frac{k_{\phi}}{\omega}\right).
\end{align}%
Notice that 
\begin{align}
 b\equiv \frac{k_{\phi}}{\omega}   
\end{align}
has the dimensions of a length, and can be seen as an impact parameter.
The light-like condition on the four-momentum
\begin{align}
	0=g^{\mu\nu}k_{\mu}k_{\nu}=&g^{tt}\omega^2-2g^{t\phi}\omega k_{\phi}+k^2= \\=&\omega^2[\alpha^2-2g^{t\phi}\alpha\frac{r}{\sqrt{-g_{tt}}}\sin\theta+g^{tt}], \label{2grado}
\end{align}%
solved for $\alpha$, gives
\begin{equation*}
	\frac{k_{\phi}}{\omega}=\alpha\frac{r}{\sqrt{-g_{tt}}}\sin\theta=\frac{g_{t\phi}}{r\sqrt{-g_{tt}}}\sin\theta+\sqrt{\frac{r^2-g_{t\phi}^2\cos^2\theta}{-r^2g_{tt}}},
\end{equation*}%
where the positive sign for the square root has been fixed taking into account that $\alpha$ is positive.\footnote{We are assuming that $\partial_t$ is always
timelike, that is, we are considering only solutions where $g_{tt}$ and $g_{\phi\phi}$ never changes in  sign.} Hence 
\begin{align}
   \frac{k_{\phi}}{\omega} &=-r\sin\theta\, \times \\ &\left[\sqrt{(\gamma^2H)^2-(\gamma^2Hv+r\Omega)^2\cos^2\theta}-(\gamma^2Hv+r\Omega)\sin\theta\right]^{-1},
\end{align}%
where we have made use of the explicit expressions of the metric components and have introduced the Lorentz gamma factor
\begin{align}
    \gamma=\frac 1{\sqrt {1-v^2}}.
\end{align}
Then, (\ref{redshift}) becomes
\begin{align}
\label{redshift3}
	1+&z= \\
	&\frac{1}{\sqrt{-H}}\left[1+\frac{r\Omega\sin\theta}{\sqrt{(\gamma^2H)^2-(\gamma^2Hv+r\Omega)^2\cos^2\theta}-(\gamma^2Hv+r\Omega)\sin\theta}\right].
\end{align}%
This is the general formula we were looking for. It explicitly gives a relation between the photon's frequency shift and the metric and, through the solution of Einstein's equations, to the density and four-velocity of matter.

\subsection{Particular cases}
Equation (\ref{redshift3}) can be evaluated in specific cases in order to infer interesting insights and physical interpretations of the quantities appearing in the metric. As a starting point, it is instructive to consider the particular case of a rigidly rotating dust. It can be obtained imposing $H(\eta)\equiv-1$ and $\Omega=\Omega_0=$ constant in \S\ref{S2}. Moreover, imposing $\Omega_0=0$, we recover the model considered by \citet{Balasin:2006cg} and \citet{crosta2020}. Most of the terms cancel out in eq.~\ref{redshift3}), leading to 
\begin{equation}
\label{z=0}
	z_{H=-1}\equiv0.
\end{equation}%
Eq.~\ref{z=0} proves that GR corrections can be indeed non negligible since, for the rigid rotating case considered here, they have exactly the same magnitude of the special relativistic terms.

We now consider more physical cases. We start with the cases of perfectly face-on  $(\sin\theta=0)$ \S\ref{FO} and edge-on $(\lambda=0)$ \S\ref{E0} galaxies. 

\subsubsection{Face-on galaxies \label{FO}}
A galaxy that is face-on with respect to the asymptotic detector corresponds to $\lambda=\pi/2$, that is, $\sin\theta=0$. This corresponds to the fact that the projection of the four-momentum of the photon along the galactic plane is negligible.
Then, eq.~\ref{redshift3} simplifies to%
\begin{equation}
	1+z(\pi/2; \phi, r, z)= \frac{1}{\sqrt{-H}}. \label{redtrasv}
\end{equation}%
The redshift depends only on the differential rotation described by $H$ and, in the general case where $H$ is not constant, it can show an observable radial profile, as discussed in section \ref{S5}.

\subsubsection{Edge-on galaxies \label{E0}}
In the case of an edge-on galaxy, we have $\lambda=0$. Suppose now a star located at $r=|b|$ along the line of sight, for which $\phi=\pm\pi/2$, where $b$ is the impact parameter, and the sign of the angle changes from the left to the right hand side. This corresponds to assume that the photon is emitted exactly in the $\partial_\phi$ direction, so that $\sin\theta=\pm1$. Substituting in eq.~\ref{redshift3}, it gives
\begin{align}
\label{cut}
	1+z(0; \pm\pi/2, r, z) =\frac{1}{\sqrt{-H}}\left[1+\frac{r\Omega}{(v\mp1)\gamma^2(-H)-r\Omega}\right]. 
\end{align}
Using now eq.~\ref{romega-chi}, it can be rewritten in the form
\begin{align}
    1+z(0; \pm\pi/2, r, z)= \frac{(\gamma\sqrt{-H})}{\gamma^2(-H) \mp r \chi} \frac{\left(1\pm v\right)}{\sqrt{1-v^2}}, \label{cut2}
\end{align}
where the upper and lower sign refers to backward and forward emission, respectively. We note that for weak field ($H\sim -1$) and negligible non-diagonal effects $\chi \rightarrow 0$, it holds $r\Omega=v$.
Notice that in the special relativistic limit ($\gamma \sqrt{-H} \rightarrow 1$ and $\chi \rightarrow 0$) $1+z(0; \pm\pi/2, r, z) \rightarrow\left(1\pm v\right)/\sqrt{1-v^2}$.

\subsubsection{Redshift of the galactic center}
\label{z est}

 Up to now, we assumed that the galaxy is purely rotating, without any bulk motion with respect to the detector. This is of course not necessary true for a real galaxy, which concurs in the Hubble expansion and can have a possible superimposed peculiar velocity dictated by local inhomogeneities of the large scale structure. All this sums-up in the measured shift of the galaxy center.
In general, the net shift of the galaxy could affect the solution by non linear effect. However, it should be a systematical effect, as long as photons arrive from the observed galaxy through almost the same path, i.e., if the angular size of the observed galaxy is sufficiently small and gravitational lensing along the line of sight is absent/negligible.
One could expect an analogous systematical shift due to the detector orbiting within the non-static axisymmetric metric of the Milky Way.

\section{Proposal of an empirical check of the GR model \label{S5}}

It is almost universally taken that gravitational effects could be neglected when measuring the frequency shift of the light coming from an external galaxy.
Under such an assumption, the measured redshift $z$ would just be comparable to the pure kinematic Doppler effect in a Minkowski space-time, corresponding to a ``special-relativistic''  velocity $\vec{v}_{SR}$ of the source, i.e., 
\begin{equation}
	1+z=\frac{1+v_{SR}^{||}}{\sqrt{1-v_{SR}^2}}.
\end{equation}%
Since $\vec{v}_{SR}$ is assumed to be directed along $\partial_\phi$, its projection along the line of sight is $v_{SR}\sin\theta$. 

Let us now compare the widely accepted special relativistic description (SR), which accounts only for the kinematic Doppler shift, to the general relativistic description (GR), which instead includes the gravitational shift effect. The expressions for the redshift are
\begin{equation}
\label{SRv}
1+z=
\begin{cases}
    \text{SR}\quad  \frac{1+v_{SR}\sin\theta}{\sqrt{1-v_{SR}^2}}, \\
    \text{GR}\quad \frac{1}{\sqrt{-H}}\left[1+\frac{r\Omega\sin\theta}{\sqrt{(\gamma^2H)^2-(\gamma^2Hv+r\Omega)^2\cos^2\theta}-(\gamma^2Hv+r\Omega)\sin\theta}\right].
\end{cases}
\end{equation}
 Here, the quantities $H$ and $\Omega$ describe the four-velocity of the emitting star $u=(-H)^{-1/2}(\partial_t+\Omega\partial_{\phi})$, so that $\Omega$ is its angular velocity as measured with respect to the Killing vectors $\partial_t, \partial_{\phi}$, and $H$ is the normalization factor. $v$ is the speed of the star as measured by the ZAMO observer, and we refer to this speed when we write the Lorentz factor $\gamma=(1-v^2)^{-1/2}$. The off-diagonal component of the metric, which is responsible for the non-Newtonian effects, is proportional to $\chi=\Omega+H\gamma^2v/r$. For further details see sections \ref{S2} and \ref{S3}.

Different galaxies corresponds to different velocity profiles $v_{SR}$, as well as different profiles of $v(r, z)$ and $H(\eta)$. Let us consider eq.~\ref{SRv} for a disc galaxy tilted by a given angle with respect to the line of sight. Along the minor axis one measures the transverse redshift as (see eq.~\ref{redtrasv}),
\begin{equation}
\label{Redperp}
1+z^{\perp}=
\begin{cases}
    \text{SR}\quad  \frac{1}{\sqrt{1-v_{SR}^2}}, \\
    \text{GR}\quad \frac{1}{\sqrt{-H}}.
\end{cases}
\end{equation}
The SR limit is again obtained by $\gamma \sqrt{-H} \rightarrow 1$ and $\chi \rightarrow 0$, which corresponds to $\frac{1}{\sqrt{-H}} \sim \gamma$. 

A measure of the redshift along the major axis, at the same location $(r,z)$, corresponds to a different value of $\phi$. Through eq.~\ref{SRv}  with $\phi=\pm\pi/2$, we find 
\begin{equation}
\label{Redtrasv}
1+z^{||}=
\begin{cases}
    \text{SR}\quad  \frac{1\pm v_{SR}\cos\lambda}{\sqrt{1-v_{SR}^2}} , \\
    \text{GR}\quad \frac{1}{\sqrt{-H}}\left[1+\frac{\pm r\Omega\cos\lambda}{\sqrt{(\gamma^2H)^2-(\gamma^2Hv+r\Omega)^2\sin^2\lambda}-(\gamma^2Hv+r\Omega)(\pm\cos\lambda)}\right].
\end{cases}
\end{equation}
We conclude that, in the SR case, the measure of the transverse frequency shift is sufficient to determine the frequency shift along the major axis since it does depend only upon $v_{SR}$. Indeed, special relativity implies
\begin{align}
(1+z^{||})=(1+z^{\perp}) \left(1\pm \cos \lambda \sqrt{1-\frac{1}{(1+z^{\perp})^2}}\right), \label{CONDITION}
\end{align}
which is not true in the general relativistic description. This is ultimately due to the presence of the additional degrees of freedom $H$ and $\chi$. Generally speaking, by neglecting gravitational effects one chooses a particular subclass of solutions, with $\gamma (-H) \rightarrow 1$ and $\chi \rightarrow 0$. 

In principle, by measuring $z^{||}(r)$ and $z^{\perp}(r)$, we would be able to derive $H(\eta)$, $v(r)$ and all the others characteristics of the solution. In particular, we would have a certain required mass $\rho$ from eq. \ref{rho}.
Therefore, the proposed comparison would provide a good indicator of the validity of the usual SR description with respect to the more fundamental general relativistic counterparts.
\subsection{Non relativistic velocities}
Let us discuss the case of the dust velocity $v \ll 1$. For non relativistic velocities, we can rewrite eq. \ref{Redperp} as 
\begin{equation}
    \text{SR}\quad 1+ z^{\perp} \approx 1+\frac{1}{2} v_{SR}^2.
\end{equation}
On the other hand, eq.~\ref{Redperp} is not affected by taking $v \rightarrow 0$, and we still have
\begin{align}
     \text{GR}\quad 1+ z^{\perp} = \frac{1}{\sqrt{-H}}.
\end{align}
In the same spirit, we can expand the redshift along the major axis as
 \begin{align}
\label{order Omega}
	1+ z^{||}=
	\begin{cases}
	    \text{SR} \quad 1\pm v_{SR}\cos\lambda+O(v_{SR}^2), \\
	    \text{GR} \quad \frac{1}{\sqrt{-H}}\left[1\pm\frac{r\Omega\cos\lambda}{-H}\right] + O(r^2\Omega^2)+O(r\Omega v)+O(v^2)
	\end{cases}
\end{align}%
We remember that when we have non negligible diagonal effects $\chi$, $r\Omega$ is different from  $v$. The SR consistency equation eq.~\ref{CONDITION} in the non relativistic velocity limit becomes 
\begin{align}
    (z^{||})^2 \approx 2 z^{\perp} \cos^2\lambda, \label{CONDITION2}
\end{align}
while still there is no analog constraint in the GR model. The measurements of $z^{\perp}$ and $z^{||}$ fix only $H$ and $r\Omega$, while the function $\chi$ is determined by the equations of motion.

\subsection{ A crude estimation of the observational test feasibility}
From eq.~\ref{Redperp} we can expect that the order of magnitude of $H+1$ is $v_{SR}^2$. Assuming close to circular orbits (i.e., $v, v_{SR}\sim v_c$) with, e.g., $v_c=10^{-3}$ (i.e., 300 km/s), we obtain
\begin{equation}
\label{order H}
	H+1=O(v^2)\sim v_c^2=10^{-6}.
\end{equation}
Since we expect that the $-H$ factors do not carry corrections at the first order, we can give the estimation of eq.\ref{SRv} as%
\begin{align}
	1+ z^{||}=
	\begin{cases}
	    \text{SR} \quad 1\pm v_{SR} \sin\theta+O(v^2), \\
	    \text{GR} \quad   1\pm r\Omega\sin\theta+O(v^2)
	\end{cases}
\end{align}%
returning $v_{SR}\approx r\Omega$ as expected.

Once expressed in the commonly used (but not strictly correct in the case in exam) velocity terms, the expected corrections are of $\sim 100$ m s$^{-1}$ for a rotational velocity of about $300$ km s$^{-1}$. The (small) magnitude of the expected corrections to the frequency shift profiles makes this novel test not straightforward to perform. 

Optical instruments potentially able to measure the corrections to the line shifts predicted by our metric are already available. For example, the ESO ESPRESSO spectrograph mounted on the VLT proved to be able to measure frequency shifts down to $\sim 10$ cm s$^{-1}$ in spectra of stars \citep{espresso}. Such an accuracy would exceed our requirements, but a few peculiarities of the proposed test complicate its feasibility. First of all, such exquisite accuracy in searches for planets, allowing for the detection of frequency shifts significantly smaller than the intrinsic broadening of the absorption lines used, is achieved convolving a mask (which models the star photosphere with the observed spectra) so that the signal from all the photosphere absorption lines observable in the spectral range (order of few thousands) contributes to the spectral shift evaluation. In the galactic scenario the number of lines that could survive the large Doppler broadening due to the (sub-dominant) dispersion in the velocity field can be significantly smaller (up to $\sim$ 10 in the best case), implying the need for an even larger signal to noise ratio (and, therefore, observation time). Furthermore, in the galactic case the integrated flux would come only from a small region of the disc (in the opposite case the characterization frequency shift profile would not be possible). An approximate back of the envelope estimate indicated that, in order to achieve a (far from sufficient) signal-to-noise ratio of 2 for a single pointing at the scale radius  of Andromeda, an observing time of $100$ hr would be required.
More efficient instrument/telescope pairs  (even with a lower spectral resolution) are therefore needed. Possible candidates are the  HIRES spectrograph on the Keck \citep[][]{hires} or the ANDES instrument to be installed on the ELT \citep[][]{andes}. We plan to quantify in a next study the observation time needed to achieve a signal-to-noise ratio sufficient to perform the proposed test in the optical band.

Although deviations from the Newtonian rotation curves associated to the baryonic components only are measurable within a few scale radii of the stellar disc, in particular for low-mass galaxies \citep[see, e.g.,][]{Strigari}, the most constraining tests come from the outer regions of the discs \citep{rubin, gentile04}, where the stellar component becomes exponentially fainter while measurable emission from the typically more extended \citep{bosma} HI disc is still visible in the 21-cm line emission. Also in this case the intrinsic line broadening is significantly larger ($\sim$ few km s$^{-1}$) than the expected shifts, requiring a high signal to noise ratios and long observation time. We plan to complement our future study on the optical tests by performing mock observations using already online instruments \citep[e.g. large single dish radio telescopes as FAST,][]{FAST2018}, as well as those planned for the future  \citep[such as, e.g. the next generation VLA,][]{ngVLA}, to constrain the observation time needed to achieve the required accuracy.


\section{Conclusions and perspectives}\label{conclusions}
We employed a general relativistic framework to describe disc galaxies. This novel description gives rise to various non-Newtonian effects\footnote{As an example, our set up does take in account the effects of gravito-electromagnetism, and, being an exact solution, it allows to account for more general contributions, too.}, which we showed could be important and non negligible in disc galaxies. 

As an important application, here we focused on frequency shift of photons from the emission to the detector. eq. \ref{SRv} makes explicit the fact that gravitational effects are, in general, not negligible and that they do need to be taken into account in interpreting radial velocity measurements. Neglecting such gravitational effects, which can be of the same order of the special relativistic Doppler effect, is equivalent to setting to zero different components of the metric and therefore, in all respects it is equivalent to selecting a particular subclass of solutions of the GR equations. This cannot be mathematically and physically justified a priori. Observations are the only judge by which we can decide which class of solutions is appropriate an physically motivated. In this sense, we propose the observational consistency check in eq.\ref{CONDITION} (or its low energy counterpart, eq.~\ref{CONDITION2} to test whether gravitational effects are actually negligible, since from a GR perspective there is no reason a priori why the conditions posed by eq.~\ref{CONDITION} and eq.~\ref{CONDITION2} should be met. 

It must be stressed that our GR galaxy model in its present form is not sufficient to fully constrain the amount of DM present in disc galaxies. Even if the less general GR model by \citet{Balasin:2006cg} managed to achieve the outstanding result of reproducing the observed properties of the MW \citep[][]{crosta2020} without the requirement of any DM, one has to refrain from invoking Occam's razor against the DM hypothesis, as multiple and independent observational arguments in its support do exist, as discussed in the introduction. If the proposed test were to result in a metric consistent with the observed distribution of baryonic matter and with the observed frequency shift and kinematics, it would not be sufficient to reject the presence of a DM halo on individual galaxies. In order to have such a stringent -GR motivated- test of the maximum amount of DM present in disc galaxies, the model needs further improvements. 
The most important improvement would be, in our opinion, the self-consistent inclusion of a close-to-spherical dark component in the metric. Only the application of such extended model to carefully selected galaxies (characterized by, e.g., dynamically cold discs and with negligible bulge components) can put statistically significant upper limits to the amount of DM allowed by observations. 

To significantly increase the sample on which to test our model, further improvements are needed. These include (but are not limited to) the inclusion of a non negligible velocity dispersion in the stars (and gas) of the disc, and of a dynamically hot stellar component modelling the bulges often present in the few central kpcs of massive disc galaxies. Through the use of the proposed GR test, we could be able to put observational upper limits of the amount of DM mass present in disc galaxies and, in turn, to put unprecedented constraints on the very nature of DM itself.




\bibliographystyle{mnras}
\bibliography{refs.bib} 




\appendix

\bsp	
\label{lastpage}
\end{document}